# A Stringent Constraint On Alternatives To A Massive Black Hole At The Center Of NGC 4258

Eyal Maoz

Harvard-Smithsonian Center for Astrophysics,
MS 51, 60 Garden Street, Cambridge, MA 02138

E-mail: maoz@cfa.harvard.edu

## ABSTRACT

There is now dynamical evidence for massive dark objects at the center of several galaxies, but suggestions that these are supermassive black holes are based only on indirect astrophysical arguments. As emphasized by Kormendy and Richstone (1995), large $M/L$ ratios and gas motions of order $\approx 10^3\,\mathrm{km\,s^{-1}}$ do not uniquely imply a massive black hole (BH), and it is possible that the central dark objects in these galaxies are massive clusters of stellar remnants, brown dwarfs, low-mass stars, or halo dark matter.

The recent unprecedented measurement of the rotation curve of maser emission sources at the center of NGC 4258, and the remarkable discovery that it is Keplerian to high precision, provides us a unique opportunity for testing alternatives to a BH. We use a conservative upper limit on the systematic deviation from a Keplerian rotation curve to constrain the mass distribution at the galaxy center. Based on evaporation and physical collision time-scale arguments, we show that a central cluster is firmly ruled out, unless the cluster consists of extremely dense objects with mass $\lesssim 0.05 M_{\odot}$ (*e.g.*, low mass BHs or elementary particles). Since both of these dynamically-allowed systems are very improbable for other astrophysical reasons, we conclude that a central dense cluster at the center of NGC 4258 is *very* improbable, thus leaving the alternative possibility of a massive BH.

We also show that the mass of the BH must be at least 98% of the mass enclosed within the inner edge of the masering disk ($3.6 \times 10^7\,M_{\odot}$). A substantial contribution to that mass from a density cusp in the background mass distribution is excluded.

## 1. INTRODUCTION

Observations provide dynamical evidence for $\sim 10^6$-$10^{9.5} M_\odot$ dark *objects* at the center of several galaxies (Kormendy & Richstone 1995, and references therein). In most cases the evidence is based on the increase of the mass-to-light ratio toward the galaxy center to values of $M/L \gtrsim 100$, and in two cases (M87, NGC 4258) it is based on gas-dynamics measurements. Suggestions that these objects are supermassive black holes rely on indirect astrophysical arguments, but as has been previously emphasized (*e.g.*, Kormendy 1993; Kormendy & Richstone 1995), large $M/L$ ratios and gas velocities of order $\approx 10^3\,\mathrm{km\,s^{-1}}$ do not uniquely imply a massive black hole (hereafter, BH). It is possible that the central dark objects in these galaxies are massive clusters of stellar remnants, brown dwarfs, low-mass stars, or halo dark matter.

A rigorous proof of a BH would be, for example, a detection of relativistic velocities at a few Schwarzschild radii, but no observation is likely to have the necessary resolution for that in the near future. Kormendy and Richstone (1995) point out that it would be an important progress if we could rule out the alternative to a BH on physical grounds. In this *Letter* we show that a dense stellar cluster at the center of NGC 4258 is essentially ruled out.

Observations of line emission from water masers near the center of the galaxy NGC 4258 (Miyoshi *et al.* 1995) have recently provided compelling evidence for rotating disk of gas surrounding a mass of $3.6 \times 10^7\,M_\odot$ within a region of 0.13 pc in radius. The existence of a circumnuclear disk in this galaxy has already been indicated by previous observations (Nakai, Inoue & Miyoshi 1993; Watson & Wallin 1994), but the case for a massive BH became strong only with the unprecedented measurement of the rotation curve of the maser sources (Miyoshi *et al.* 1995), and the remarkable discovery that it is Keplerian to high precision.

The unprecedented quality of information on the dynamics at the center of NGC 4258 provides a unique opportunity for testing the massive dark cluster hypothesis. In §2 we derive a severe constraint on the nature of a central cluster, based on the *accuracy* of the Keplerian fit to the rotation curve of the masering disk. In §3 we investigate the possibility that the mass enclosed within the inner edge of the disk is in the form of a relatively light BH, surrounded by an enhanced mass density field of bulge stars or dark matter. In §4 we summarize the conclusions.

## 2. A DENSE STELLAR SYSTEM

The high-velocity maser emission features in NGC 4258 are offset from the center of rotation in a nearly planar structure, and their velocity decreases from $v_{in}$ ($1080\pm2\ \mathrm{km\,s^{-1}}$) at a distance $r_{in}$ (0.13 pc), to $v_{out}$ ($770\pm2\ \mathrm{km\,s^{-1}}$) at a distance $r_{out}$ (0.25 pc). Deriving the rotation curve of the masering disk under the naive assumptions of a perfectly planar disk and exact circular motions reveals that it can be fit by a Keplerian relation remarkably well (Miyoshi *et al.* 1995). The upper limit for a systematic deviation of the velocity profile from a Keplerian relation is $\Delta v \lesssim 3\,\mathrm{km\,s^{-1}}$ (Jim Herrnstein & Jim Moran, private communication), that is, a fractional deviation of $\Delta v/\bar{v} \lesssim 4 \times 10^{-3}$, where $\bar{v}$ is the average rotational velocity in the masering disk (notice that an uncertainty in the galaxy distance affects the mass determination, but not the accuracy of the Keplerian fit).

This is a truly remarkable result, especially in view of the many conceivable ways to introduce deviations from an exact Keplerian rotation curve: the gravitational potential field due to the mass distribution of stars, dark matter and gas in this region, non-gravitational forces on the gas, a warp in the masering disk, non-circular motions, etc. In particular, we notice that the presence of a dense stellar cluster at the galaxy center, rather than a single point mass, would also contribute to a systematic deviation from a Keplerian rotation curve if the mass distribution of the cluster extends beyond the inner edge of the masering disk ($r_{in}$). Since all the above sources for non-Keplerian deviations are unlikely to conspire together in producing an accurate Keplerian fit, it is safe to conclude that the systematic deviation from a Keplerian rotation curve due to an hypothetical central cluster alone does not exceed the deviation caused by all possible effects combined ($\Delta v \lesssim 3\,\mathrm{km\,s^{-1}}$).

Let us assume that at the center of NGC 4258 resides a dense spherically symmetric cluster, where the cluster mass enclosed within a sphere of radius $r_{in}$ is $M(<r_{in}) = 3.6 \times 10^7\,M_\odot$. Denoting the ratio of the cluster mass enclosed between the spheres of radii $r_{in}$ and $r_{out}$ to its mass within $r_{in}$ by $\delta_{Kep} \equiv [M(<r_{out}) - M(<r_{in})]\,/M(<r_{in})$, it is easy to show that in the limit $\delta_{Kep} \ll 1$ we have $\delta_{Kep} = 2\Delta v/\bar{v}$, where $\bar{v} \equiv (v_{in} v_{out})^{1/2}$. In light of the above discussion and the current observational upper limit on $\Delta v/\bar{v}$ ($\lesssim 0.004$) we conclude that a conservative upper limit on $\delta_{Kep}$ would be $\delta_{Kep} \lesssim 1\%$.

Let us consider the hypothesis that the entire mass within a radius $r_{in}$ is in a dense cluster with a mass density profile of the general form $\rho_0\,[1 + (r/r_c)^2]^{-\alpha/2}$. If $\alpha = 2$ the cluster mass increases with radius at least as fast as $r$, regardless of the core radius, and we obtain $\delta_{Kep} \gtrsim 1$. Clearly, an $r^{-2}$ density falloff is far from being steep enough to avoid contributing too much mass within $[r_{in}, r_{out}]$. The mass density profile of elliptical galaxies has an asymptotic logarithmic slope of $\simeq -4$, and that of globular clusters is $\simeq -5$ (the steepest density profile observed in astrophysical systems). Assuming $\alpha = 5$ (a Plummer model) we have

$$\rho(r) = \rho_0 \left(1 + \frac{r^2}{r_c^2}\right)^{-5/2} \quad ; \quad M(<r) = \frac{4\pi\rho_0}{3} r^3 \left(1 + \frac{r^2}{r_c^2}\right)^{-3/2} \quad . \tag{1}$$

Requiring that $[M(<r_{out}) - M(<r_{in})]/M(<r_{in}) \lesssim \delta_{Kep}$, we obtain a constraint on the core radius of the cluster

$$r_c \lesssim \left[(1 + \delta_{Kep})^{2/3} - 1\right]^{1/2} \left[\frac{1}{r_{in}^2} - \frac{(1 + \delta_{Kep})^{2/3}}{r_{out}^2}\right]^{-1/2} \quad . \tag{2}$$

Substituting $\delta_{Kep} = 0.01$ yields $r_c \lesssim 0.012$ pc. Combined with the requirement that $M(<0.13\,\mathrm{pc}) = 3.6 \times 10^7\,M_\odot$, the central mass density is $\rho_0 \gtrsim 4.5 \times 10^{12} M_\odot\,\mathrm{pc}^{-3}$. The central velocity dispersion in a Plummer model is given by $\sigma_0 = [(2\pi/9) G \rho_0 r_c^2]^{1/2}$, and we obtain $\sigma_0 \gtrsim 1500\,\mathrm{km\,s^{-1}}$.

### 2.1. The Evaporation Time-Scale

An upper limit to the lifetime of any bound stellar system is given by its evaporation time. The evaporation time-scale of clusters of objects with a single mass (Spitzer & Thuan 1972; Binney & Tremaine 1987, hereafter BT) is $t_{evap}\approx 300\,t_{relax}$, where $t_{relax}$ is the median relaxation time which is given by $t_{relax}=[0.14N/\log(0.4N)]\,(R_{1/2}^3/GM_{cl})^{1/2}$ (Spitzer & Hart 1971; BT), $N$ is the number of objects in the cluster, $M_{cl}$ is the cluster mass, and $R_{1/2}$ is the cluster radius within which lies half of the cluster's mass. In the case of a Plummer model $R_{1/2}=1.3\,r_c$, and since $r_{in}\gg r_c$ we have $M_{cl}\simeq 3.6\times 10^7\,M_\odot$. Assuming that the hypothetical cluster at the center of NGC 4258 consists, for example, of neutron stars ($m=1.4M_\odot$) we obtain

$$t_{evap}\lesssim 3.4\times 10^8\ \mathrm{yr} \quad . \tag{3}$$

Since this period of time is much shorter than the age of the galaxy, a cluster of objects with mass $\approx 1.4M_\odot$ at the center of NGC 4258 is firmly ruled out. The lower the objects' typical mass, the longer is the evaporation time-scale. A similar cluster which consists of objects with mass of $0.05M_\odot$ would evaporate within $t_{evap}\lesssim 8\,\mathrm{Gyr}$.

We argue that it is impossible to increase significantly the evaporation time-scale by reasonable modifications of the above derivation. First, one should notice that all the inequality signs in the above discussion result from the constraint $\delta_{Kep}\lesssim 10^{-2}$. If the systematic deviation from a Keplerian rotation curve had been even as twice as large, the evaporation time would have gone up only by a similar factor. If the density profile of the cluster was flatter (*e.g.*, $\alpha=4$), the core radius would have to be $\lesssim 0.003$ pc in order to be consistent with the constraint on $\delta_{Kep}$, and this would result in an outrageous central mass density of above $10^{14}M_\odot\,\mathrm{pc}^{-3}$ and consequently an even stronger constraint on the evaporation time-scale. If the density profile was much steeper (*e.g.*, $\alpha=7$, although unreasonable) we would get $r_c(\alpha=7)=3\,r_c(\alpha=5)$, and $t_{evap}(\alpha=7)=2.3\,t_{evap}(\alpha=5)$, which would not make a qualitative difference. The fact that the cluster is not isolated but is embedded within a galaxy is insignificant, since the depth of the cluster potential well (proportional to $\sigma^2$) is about two orders of magnitude larger than that of the surrounding galaxy. Small deviations from spherical symmetry in the structure of the cluster will introduce changes of order unity. We ignored the mass of the masering gas enclosed within $[r_{in},r_{out}]$ ($\lesssim 2\times 10^5 M_\odot$ for avoiding thermal quenching of the maser [Miyoshi *et al.* 1995]). If we took it into account, it would imply an even stronger constraint on the cluster mass enclosed within $[r_{in},r_{out}]$ (in order to satisfy $\delta_{Kep}\lesssim 1\%$), and this would make the constraints on $r_c$ and $\rho_0$ stronger.

We conclude that, *based on the evaporation time-scale argument, a dense stellar system which consists of objects with typical mass* $\gtrsim 5\times 10^{-2}\,M_\odot$ *at the center of NGC 4258 is ruled out.* As we shall now see, a cluster of objects with lower masses is very improbable due to a different constraint.

## 2.2. The Physical Collision Time-Scale

The characteristic time-scale in which a star suffers a physical collision is given by (BT)

$$t_{coll} = \left[16\pi^{1/2} n \sigma r_*^2 \left(1 + \frac{Gm}{2\sigma^2 r_*}\right)\right]^{-1} \quad , \tag{4}$$

where $n$ is the number density of stars, $r_*$ is the radius of the star, $m$ is the mass of the star, and $\sigma$ is the velocity dispersion. The first term in the brackets is simply due to geometry, and the second term represents the enhancement in the collision rate by gravitational focusing. The mass-radius relation for zero-temperature brown dwarfs and very low-mass objects of cosmic composition is well fit by

$$r_* \;=\; 2.2 \times 10^9 \left(\frac{m}{M_\odot}\right)^{-1/3} \left[1 + \left(\frac{m}{0.0032 M_\odot}\right)^{-1/2}\right]^{-4/3} \quad \mathrm{cm} \tag{5}$$

(Zapolsky & Salpeter 1969; Stevenson 1991). The radius scales as $m^{1/3}$ for very small masses, where the gravitational forces are small compared with the electrostatic forces, and as $m^{-1/3}$ for large masses, where the electrostatic forces are small compared with the Fermi pressure and with gravitation. The maximum radius is reached between these two regimes at a mass of $3.2\times10^{-3} M_\odot$ (see detailed discussions in the papers mentioned above).

First, it is easy to verify that the gravitational focusing term in equation (4) is negligible for any $m \lesssim 0.05 M_\odot$ (§2.1) and $\sigma \gtrsim 1500\ \mathrm{km\,s^{-1}}$ (§2). The larger the stellar radius, the lower is the collision time. Since the radius of a finite-temperature object is larger than that given by equation (5) (*e.g.*, Stevenson 1991), substituting equation (5) in equation (4) will give an upper limit to the collision time-scale. Taking cluster parameters as derived following equation (2), $n \approx \rho_c/m$, where $\rho_c$ is the average mass density within the core of the Plummer model, we find that the time-scale for physical collisions in the hypothetical cluster at the center of NGC 4258 must be below the solid curve in Figure 1.

We see that, regardless of the stellar mass, the collision time-scale is an extremely short period of time relative to the age of the galaxy. Since physical collisions may lead to coalescence of the two stars, and would certainly cause a rapid evolution of the stellar mass function (roughly doubling the typical stellar mass within successive periods of time $t_{coll}[m]$), we conclude that *a cluster which consists of objects with typical mass* $\lesssim 0.05 M_\odot$ *(the range of masses allowed by the constraint on the evaporation time-scale) is ruled out by the constraint on the physical collision time-scale, unless the objects are extremely dense (e.g., light black holes or elementary particles)*. It is straightforward to see that if the mean density of the low-mass objects exceeds that given by equation (5) even by a couple of orders of magnitude, it would not make a significant qualitative difference[1].

---

[1] Similar time-scale considerations were applied to the mass distribution near the centers of M31 and M32 (Richstone, Bower & Dressler 1990), but present observations provide only weak constraints on the existence and nature of massive dark object in these systems.

## 3. A LOW MASS BH SURROUNDED BY A DENSITY CUSP?

Could the detected mass within the inner edge of the masering disk be in the form of a relatively light BH, surrounded by a massive density cusp in the background mass distribution (stars and dark matter)? We shall now show that this is not a viable possibility and that the BH mass must be very close to $3.6\times10^7\,M_\odot$.

Any massive BH at the galaxy center is inevitably surrounded by a dense stellar distribution. If the mass density field near the BH can be fit by a power law $\rho(r)=\rho'(r'/r)^\alpha$, the background mass enclosed within a radius $r$ is

$$M(<r) = \frac{4\pi\rho' r'^\alpha r^{3-\alpha}}{3-\alpha} \quad . \tag{6}$$

As in the case of a dense stellar cluster (§2), the upper limit on the systematic deviation from a Keplerian rotation curve implies that the ratio of the background mass enclosed between the spheres of radii $r_{in}$ and $r_{out}$, to the mass enclosed within a radius $r_{in}$ should not exceed $\delta_{Kep}$. This constraint now reads

$$M(<r_{out}) - M(<r_{in}) \lesssim \delta_{Kep}\left[M_{BH} + M(<r_{in})\right] \quad . \tag{7}$$

Since equation (6) implies that $M(<r_{out}) = (r_{out}/r_{in})^{3-\alpha} M(<r_{in})$, we obtain from equation (7)

$$\frac{M(<r_{in})}{M_{BH}+M(<r_{in})} \lesssim \delta_{Kep}\left[\left(\frac{r_{out}}{r_{in}}\right)^{3-\alpha} - 1\right]^{-1} \quad . \tag{8}$$

Bahcall and Wolf (1976) used Fokker-Planck analysis to derive the mass distribution of stars around a massive BH under the assumptions of a spherically symmetric background and an isotropic velocity distribution. They found that the equilibrium star mass density around the BH is given by

$$\rho(r) \approx \rho'\left[1 + \left(\frac{r_h}{r}\right)^{7/4}\right] \quad , \tag{9}$$

where $\rho'$ is the unperturbed background mass density, and $r_h$ is the characteristic gravitational capture radius (or the radius of the "sphere of influence"), which is given by

$$r_h \equiv \frac{GM_{BH}}{\sigma^2} = 15\left(\frac{M_{BH}}{3.6\times10^7\,M_\odot}\right)\left(\frac{\sigma}{100\,\mathrm{km\,s^{-1}}}\right)^{-2}\ \mathrm{pc}\,. \tag{10}$$

Assuming that the unperturbed central mass density of the bulge was uniform on the scale of $r_h$, equation (9) implies that $\rho(r \ll r_h) \propto r^{-7/4}$. Peebles (1972) used a different approach to investigate the same problem and concluded that $\rho(r \ll r_h) \propto r^{-9/4}$. Binney & Tremaine (1987) have shown that the density cusp of stars which are bound to the hole is $\propto r^{-3/2}$ when the age of the central mass is small compared to the relaxation time. When the age of the system is much larger than the relaxation time, they obtained $\rho \propto r^{-7/4}$ as did Bahcall and Wolf (1976), and this has been verified by numerical solutions of the Fokker-Planck equation (Cohn & Kulsrud 1978, and references therein). These theoretical predictions are reasonably consistent with observations. For example, the density profile at the center of M87 has a logarithmic slope of $\simeq -5/4$ (Young *et al.* 1978; Lauer *et al.* 1992), as inferred from the surface brightness profile, and that of the star distribution near the Galactic center is $\simeq -1.8$ (Bailey 1980; Genzel & Townes 1987, and references therein).

In light of these results we may assume that the power law index of the density cusp at the center of NGC 4258 is $5/4 \lesssim \alpha \lesssim 9/4$, in which case the inverse of the square brackets on the r.h.s of equation (8) is between 0.5 and 1.6. Thus, an upper limit to the background mass within the inner edge of the disk, relative to the total mass within this region ($3.6\times10^7\,M_\odot$), is

$$\frac{M(<r_{in})}{M_{BH}+M(<r_{in})} \lesssim 0.016\left(\frac{\delta_{Kep}}{1\%}\right) \quad . \tag{11}$$

The above derivation assumes that the BH mass is large enough so that $r_h$, as defined in equation (10), is much larger than the scale of the masering disk. If the BH mass had been small enough ($M_{BH} \lesssim 3\times10^5 M_\odot$) such that $r_h \lesssim r_{in}$, the background density field within the region of the masering disk would have been roughly uniform (see equation [9]), thus implying $\delta_{Kep} > 1$. Therefore, we conclude from equation (11) that *the BH mass must be at least* 98% *of the mass inferred by the Keplerian fit to the rotation curve of the masering disk.*

## 4. CONCLUSIONS

The rotation curve of the maser emission sources at the center of NGC 4258 is remarkably Keplerian (a conservative upper limit on the dimensionless deviation from a Keplerian relation is $\lesssim 10^{-2}$). This observation provides a strict constraint on the mass distribution within $\lesssim 0.25$ pc from the center of the galaxy nucleus, thus constraining the characteristics of an hypothetical central massive cluster.

Based on an evaporation time-scale argument, we have shown that a dense stellar system at the center of NGC 4258 which consists of objects with typical mass above $5\times10^{-2}\,M_\odot$ is ruled out. Lower mass objects, such as brown dwarfs and very-low mass objects of cosmic composition, are ruled out by a constraint on the physical collision time-scale in the cluster.

A central massive cluster of extremely dense objects with mass $\lesssim 0.05\,M_\odot$ (*e.g.*, low-mass BHs or elementary particles) is plausible but very improbable: a cluster which consists of elementary particles is not excluded by the dynamical arguments discussed here, but it would be a serious challenge for theory to explain how such dense compact systems of collisionless elementary particles could have formed at the center of galactic nuclei. A dense cluster of low mass ($\lesssim 0.05 M_\odot$) black holes is also dynamically allowed (the time-scale for energy dissipation due to gravitational radiation is extremely long), but we are not aware of any theory of structure formation or stellar evolution which may suggest the possible existence of a large population of such low-mass BHs.

Upcoming repeated observations of the maser emission in this galaxy may enable to pin down the structure of the masering disk even better (based on the expected proper motions of the maser sources), which may provide a lower upper-limit to the deviation from a Keplerian fit, and thus an even stronger constraint on the hypothetical cluster. But already now, the case for a massive BH at the center of NGC 4258 is very strong. As shown in §3, the BH mass must be at least 98% of the mass inferred by the Keplerian fit to the rotation curve of the masering disk ($3.6 \times 10^7 M_\odot$).

I thank George Field for many stimulating discussions and valuable comments, and Jim Moran, Jim Herrnstein, Lincoln Greenhill and Charles Gammie for discussions. This work was supported by the U.S. National Science Foundation, grant PHY-91-06678.

This preprint was prepared with the AAS LaTeX macros v3.0.

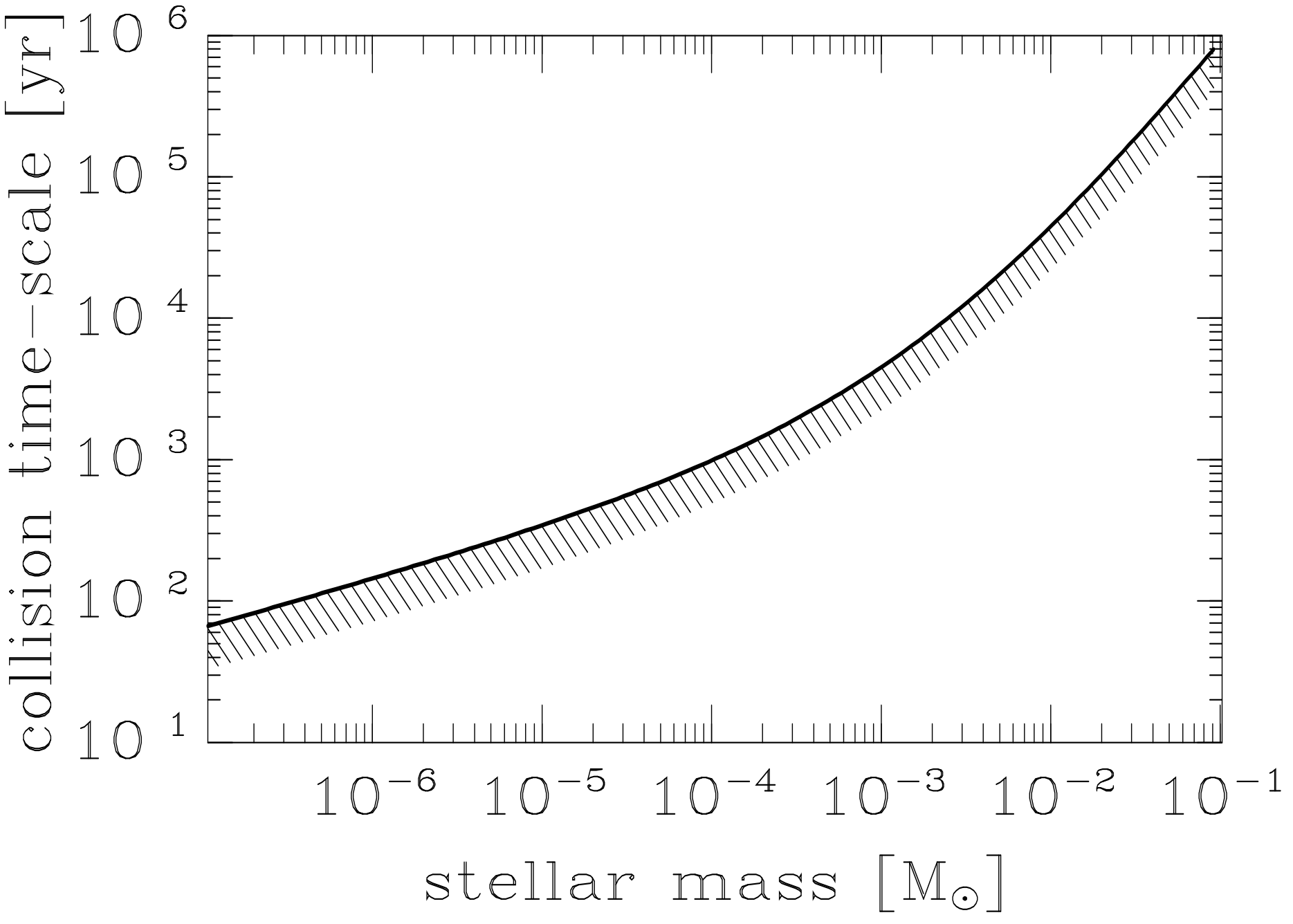

Fig. 1.— The region below the solid curve gives the time-scale for physical collisions in the hypothetical massive cluster at the center of NGC 4258, if it consists of low-mass stars, brown dwarfs, or very low-mass objects of cosmic composition (see text). The mass range $m \gtrsim 0.05\,M_\odot$ is ruled out by an evaporation time-scale argument (§2.1).